%% file: main.tex
\documentclass{IEEEtran}
\usepackage{graphicx} 
\usepackage{orcidlink}
\usepackage{subcaption}
\usepackage{amsmath,amssymb,amsfonts}
\usepackage{adjustbox}

\usepackage{xfrac}
\usepackage{textcomp,nicefrac}
\usepackage[switch]{lineno}
\usepackage[inkscapelatex=false]{svg}
\usetikzlibrary{dsp,chains}
\DeclareMathAlphabet{\mathpzc}{OT1}{pzc}{m}{it}

\newcommand{\chebyu}[2]{U_{#1}\left(#2\right)}
\newcommand{\ssrsample}[2]{#1\left[#2\right]}
\def\BibTeX{{\rm B\kern-.05em{\sc i\kern-.025em b}\kern-.08em
T\kern-.1667em\lower.7ex\hbox{E}\kern-.125emX}}
\begin{document}

\title{Trigger system for the Payload for Ultrahigh Energy Observations (PUEO) balloon-borne neutrino detector}

\include{pueoieee_authors.tex}
\renewcommand\footnotemark{}
\renewcommand\footnoterule{}
\thanks{Manuscript submitted July 7, 2026. See the acknowledgements sections for support information.}
\include{pueoieee_institutes.tex}

\bstctlcite{IEEEexample:BSTcontrol}

\maketitle

\begin{abstract}
The Payload for Ultrahigh Energy Observations (PUEO) is a NASA balloon-borne
instrument for the detection of ultra-high energy (UHE) neutrinos with
energies above $10^{17.5}~\textrm{eV}$ via either the Askaryan effect or geomagnetic
emissions from an upward-going air shower. The main instrument trigger system
for PUEO is a fully digital supersample rate beamformer based on 24 Xilinx Radio
Frequency System-on-a-Chip (RFSoC) digitizers sampling 192 channels operating 
at $3~\textrm{GSa/s}$ and a system clock frequency of $375~\textrm{MHz}$. The
trigger implements frequency band conditioning, dynamic radio-frequency
interference (RFI) rejection, and matched filtering, with significant emphasis
on optimization to reduce both the power and resource usage while maintaining
sensitivity. The system implements 48 total synthetic antenna beams with
up to 8 antennas each, covering a $\sim25^\circ$ range in zenith and
$\sim60^\circ$ range in azimuth. Preflight testing demonstrated a trigger
performance of a minimum signal-to-noise ratio (SNR) of $\sim1.5$ using
simulated signals while consuming between $5-7~\textrm{W}$ in the trigger logic.
\end{abstract}


\section{Introduction}

At ultra-high energies (UHE, $>10^{17.5}~\textrm{eV}$), neutrino interactions produce
broadband, highly impulsive radio emission from high-energy cascades either in
the ice through the Askaryan effect \cite{PhysRevLett.99.171101} or via
geomagnetic emission \cite{PhysRevLett.116.141103}. The Payload for Ultrahigh
Energy Observations (PUEO) \cite{Abarr_2021} is a balloon-borne instrument designed for the detection of these neutrinos above the Antarctic ice. It represents a
significant improvement in balloon-borne detection of radio signals from UHE particle cascades either via the Askaryan effect or
geomagnetic emissions from upward-going air showers. PUEO utilizes direct 
radio-frequency (RF) digitization at $3~\textrm{GSa/s}$ to coherently sum the
signals from up to 8 antennas in multiple (48 implemented) beamformed
directions at once. In addition, significant pre-trigger signal processing is
used to enhance the overall signal-to-noise ratio (SNR) prior to the
beamform process. Here we cover the design, implementation, and preflight
performance of the trigger system for PUEO. Full performance of the trigger
during the 2025-2026 flight will be covered elsewhere.

This work expands on the ARA experiment's phased array detector \cite{ALLISON2019112} 
and the balloon-borne ANITA UHE neutrino detector \cite{PhysRevD.99.122001}. The main
contributions of this work are the implementation of a fully digital trigger
chain at $3~\textrm{GSa/s}$, a low-power field-programmable gate array (FPGA) realization of a matched filtering and
beamforming, a resource-efficient radio-frequency interference (RFI) mitigation
implementation, and preflight validation of the resulting trigger efficiency.

\subsection{Instrument overview}

The PUEO main instrument consists of 24 Sampling Unit for RF (SURFv6) modules
with Trenz Elektronik TE0835 modules, each containing a Xilinx Radio Frequency 
System-on-a-Chip (RFSoC) XCZU47DR, a combination of a high-density programmable logic (PL) fabric and a quad-core ARM processing system (PS),
and an 8-channel high-speed ADC configured to operate at $3000~\textrm{MSa/s}$.
Each SURF handles 8 individual RF channels, giving a total of 192 input channels, 
split between horizontal and vertical polarizations. An additional two SURFs handle a
separate low-frequency instrument not described here.

Each SURF's 8 RF channels consist of two ``phi sectors'', an azimuthal slice
of $15^{\circ}$ containing four antennas sharing the same boresight direction.
One antenna in each phi sector is located in the top ring, with a baseline
distance to the highest antenna in the bottom rings of either $3.82~\mathrm{m}$
or $3.25~\mathrm{m}$, with the remaining three antennas in the bottom ring
spaced by $0.725~\mathrm{m}$. 

\begin{figure}
\begin{subfigure}{\linewidth}
\begin{center}
\includegraphics[width=0.5\linewidth]{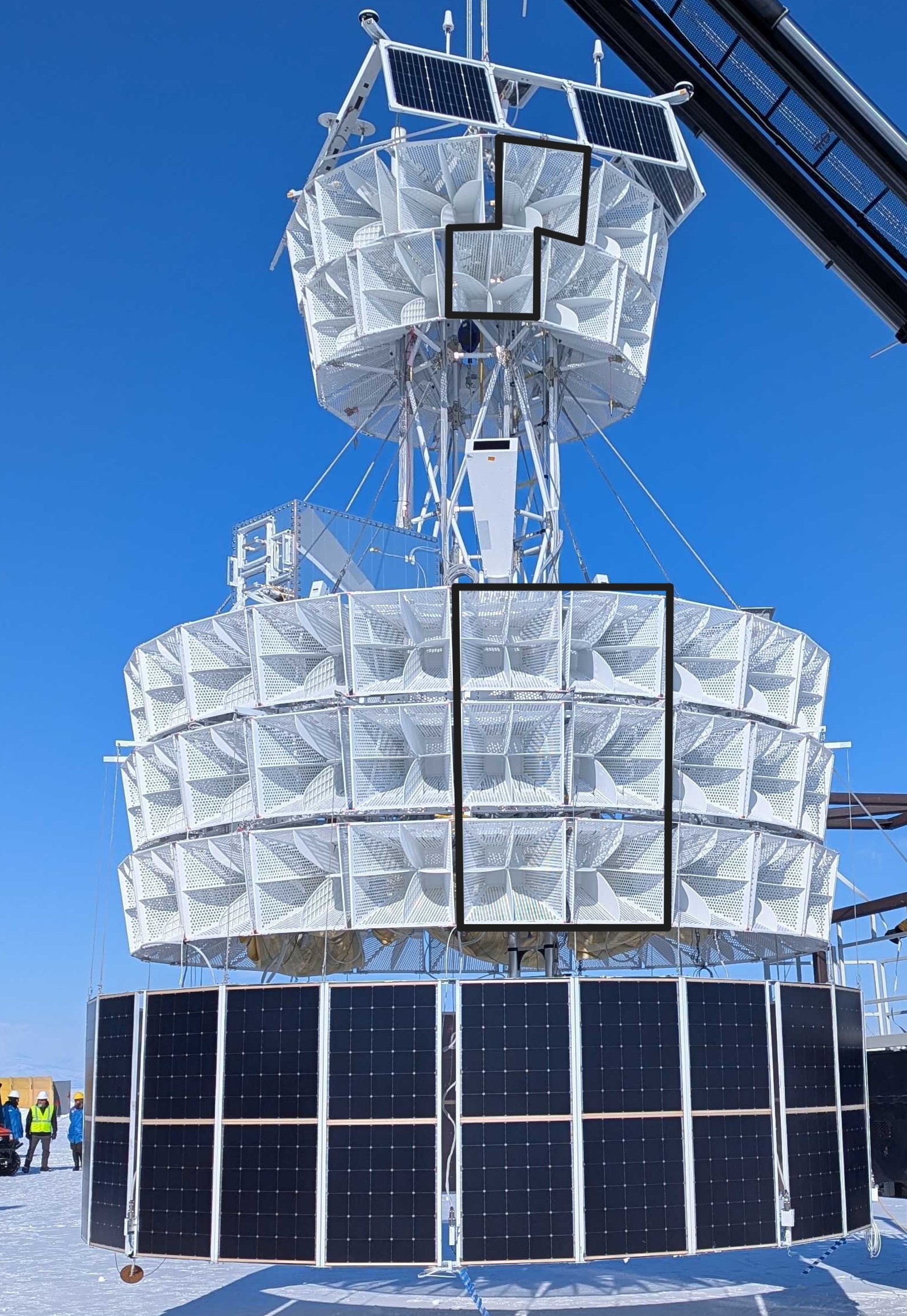}
\end{center}
\caption{PUEO payload}
\label{fig:payload}
\end{subfigure}
\begin{subfigure}{\linewidth}
\begin{center}
\includegraphics[width=0.5\textwidth]{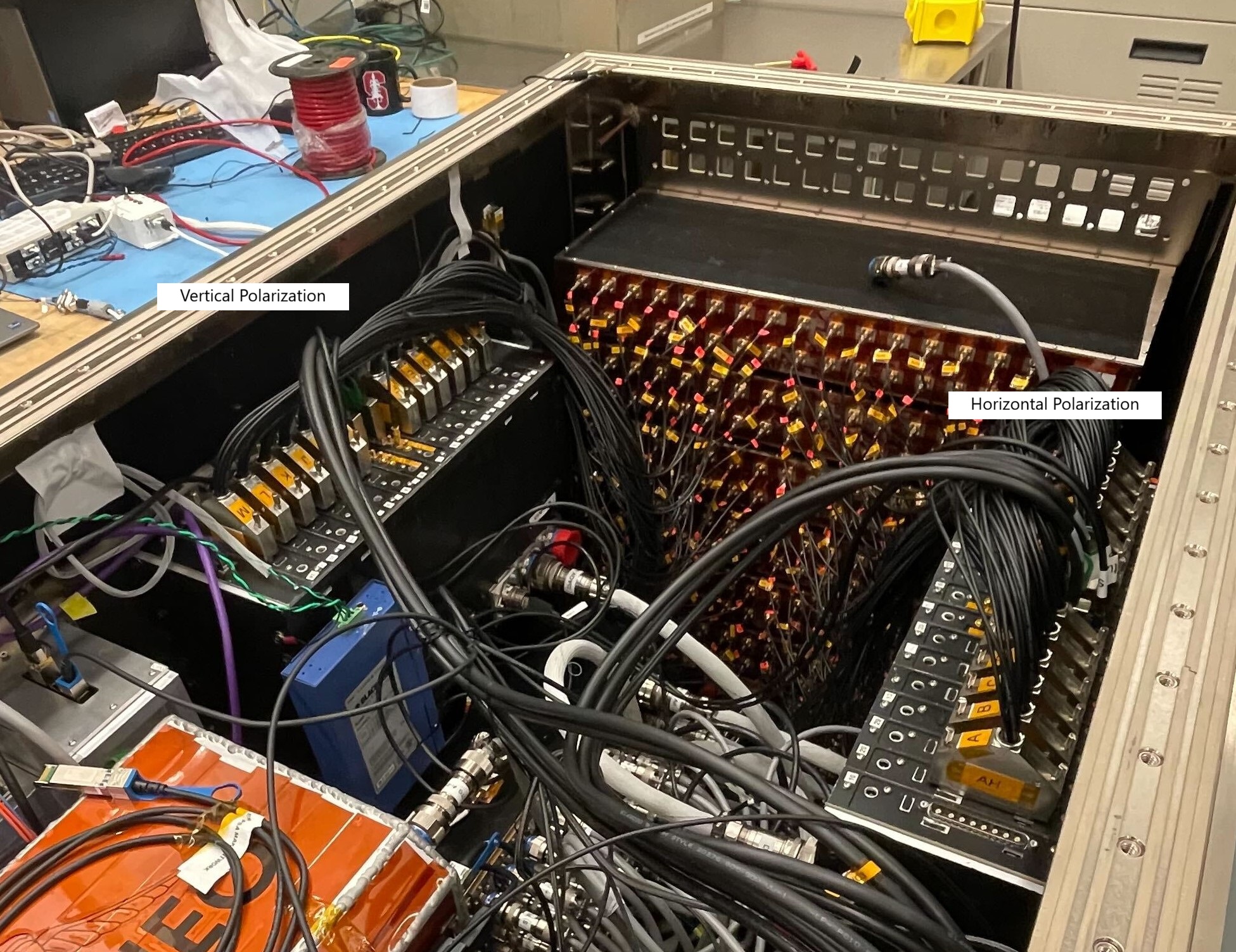}
\end{center}
\caption{Main instrument enclosure}
\label{fig:mie}
\end{subfigure}
\caption{The PUEO payload and main instrument enclosure. The outlined antennas
indicate the structure of the 8 antennas fed into each SURF, consisting of 2
4-antenna ``phi sectors'' viewing a $15^{\circ}$ slice of azimuth. Each antenna
has both horizontal and vertical polarizations, recorded in separate data
acquisition crates as shown.}
\label{fig:payloadandmie}
\end{figure}

The antennas themselves are quad-ridged horn antennas designed by Toyon Research
Corporation specifically for PUEO, sensitive to the $300-1200~\mathrm{MHz}$ frequency
range. The antennas are dual polarization, with the horizontal and vertical
polarizations each independently recorded and triggered by two separate data
acquisition crates (shown in Fig.~\ref{fig:mie}). The PUEO instrument is shown in
Fig.~\ref{fig:payload} indicating the 8 antennas which feed into each SURF.
The antennas have a very stable off-axis response, allowing for beams to be
formed between phi sectors without amplitude and phase correction over frequency.

\subsection{Architecture}

The trigger is split into multiple levels, with individual beam triggers merged into
level one (L1) triggers at each SURF, and multiple L1s combined into a level two (L2)
trigger, which are then combined to generate a global experiment trigger at the
Trigger Unit for RF (TURFv6). The L1 implementation is purely in the PL of each SURF,
with the PS used only at initialization and powered down in operation for power
and thermal purposes. The L2 is located in the programmable logic at the TURF.

Inside the SURF, data from the RFSoC is immediately stored unaltered into a large ring
buffer (synchronized between all SURFs) corresponding to $32.768~\mu \textrm{s}$ 
($98304~\textrm{samples}$), which allows for a large trigger latency. The trigger
itself is transmitted as a 12-bit timestamp along with additional metadata, 
which is then translated into a readout address in the buffer. This results in a
trigger time quantization of $24~\textrm{samples}$. The position of the trigger in
the $1024~\textrm{sample}$ readout window is programmable and was tuned to place
the trigger approximately at $300~\textrm{samples}$ ($100~\textrm{ns}$).

\begin{figure}
\begin{center}
\includegraphics[width=0.9\linewidth]{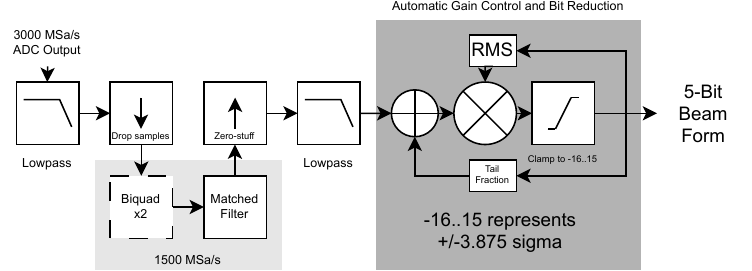}    
\end{center}
\caption{Block diagram of the trigger signal processing chain.
The optional biquad block and the matched filter operate at
$1.5~\textrm{GSa/s}$ and the AGC-BR block operates at
$3~\textrm{GSa/s}$.
}
\label{fig:triggerfilterchain}
\end{figure}

The PUEO trigger is fully digital and is discussed in three stages, with an
overall block diagram shown in Fig.~\ref{fig:triggerfilterchain}. The first stage
is a filter chain that shapes the signal response to improve SNR and optionally 
suppress man-made RF interference. The second stage is an automatic
gain control and bit-reduction block (AGC-BR) which normalizes channel amplitudes
while reducing resource usage in the beamformer. The final stage performs beamforming
and envelope-based thresholding to generate the L1 trigger.

\section{Trigger filter chain}

The filter chain consists of (in order) a downsampling block, a matched filter,
an optional RFI rejection block, and finally an upsampling block before the
AGC-BR. The RFI rejection portion of the filter chain consists of up to 2
programmable digital biquad filters. Due to power management needs, builds
containing 0, 1, or 2 biquads are selectable during normal operation. Both
the upsampling and downsampling blocks are based on the same 31-tap finite
impulse response halfband filter.

The downsampling, upsampling, and AGC-BR blocks operate at 
$3000\,\text{MSa/s}$ with a supersample rate (SSR) factor of 8 ($M=8$).
The intermediate (matched filter, RFI rejection) blocks operate at
$1500\,\text{MSa/s}$ with an SSR factor of 4 ($M=4$), both at a system
clock rate of $375\,\text{MHz}$.

The output of each intermediate block is rounded and constrained to an
overall signed 12-bit range. The final AGC-BR block scales the inputs to
a common scale and reduces the total range to 5 bits (32 total values).

\subsection{Downsample/upsample blocks}

While significant power from the Askaryan signal can be observed throughout the
band, the overall SNR peaks in the lower half for the majority of signals.
This is due to both the off-axis gain profile of the antennas
as well as the Askaryan signal frequency spectrum. We therefore use a 
low-pass (halfband) filter to reduce the overall bandwidth to below
$750\,\text{MHz}$, which also allows the remainder of the trigger chain 
to decimate by a factor of 2 and operate at $1500\,\text{MSa/s}$,
reducing power consumption. However, before the AGC-BR block, the signal is restored
to $3000\,\text{MSa/s}$ by zero stuffing followed by the same halfband 
filter, allowing for finer beamforming delays.

The halfband filter is a 31-tap FIR filter operating at an $M=8$. The transfer 
function in the $z-\textrm{domain}$ (with $z^{-1}$ the discrete unit time delay), 
organized to show the SSR structure, is:


\[
\resizebox{\hsize}{!}{%
	$K^{-1}H(z)=
	\begin{matrix}
		 & -23z^{-1} &  & +105z^{-3} &  & -263z^{-5} &  & +526z^{-7}\\
		 & -949z^{-9} &  & +1672z^{-11} &  & -3216z^{-13} &  & +10342z^{-15}\\
		+2^{14}z^{-16} & +10342z^{-17} &  & -3216z^{-19} &  & +1672z^{-21} &  & -949z^{-23}\\
		 & +526z^{-25} &  & -263z^{-27} &  & +105z^{-29} &  & -23z^{-31} \\
	\end{matrix}
	$%
}
\]

\begin{figure}
\begin{subfigure}{\linewidth}
\includegraphics[width=0.9\linewidth]{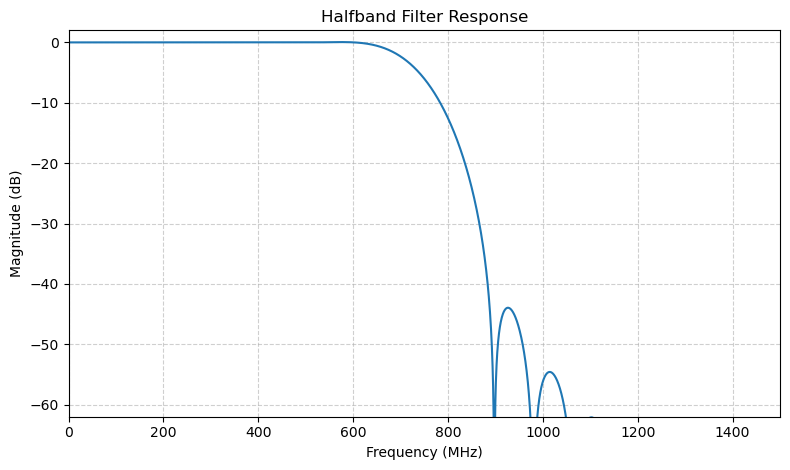}
\caption{Frequency response}
\label{fig:halfbandFreqResp}
\end{subfigure}
\begin{subfigure}{0.9\linewidth}
\begin{center}
\includegraphics{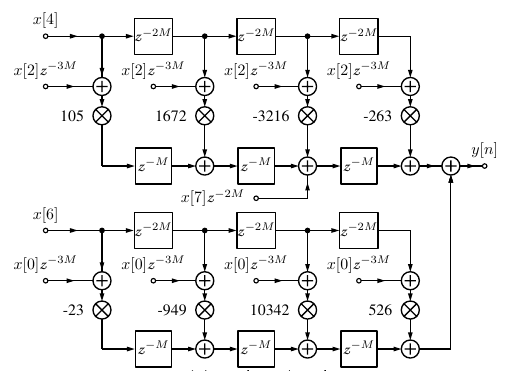}
\end{center}
\caption{Filter structure}
\label{fig:halfbandStructure}
\end{subfigure}
\caption{Halfband filter response and structure, omitting pipeline
registers. Only the structure for sample 7 is shown -- the others are
identical except for index rotation and input delays.}
\label{fig:halfbandFilter}
\end{figure}

Here $K=2^{-15}$ for the downsample block and $K=2^{-14}$ for the
upsample block, resulting in a net unity gain for the combination of the
two. The filter contains an additional unit delay to align the center tap
with the original sample, giving an overall group delay of $z^{-16}$
(two system clocks). The symmetric nature of the filter lends itself to being
organized as the sum of two 4-tap systolic filters on individual samples,
with the preadd feature of the FPGA digital signal processor (DSP) block 
used to combine the samples with common coefficients but reversed order, 
as shown in Fig.~\ref{fig:halfbandStructure}. The notation $x[i]=x z^{i}$ 
has been chosen to represent calculations in supersample rate; note that
system clocks generate a delay of $z^{-M}$, where $M=8$ in this case.

As an example, ignoring pipeline registers, for one of the filters for
$y[7]$ (so $yz^7$), the first DSP takes in $x[4]$ and preadds $x[2]z^{-3M}$. This
generates $105\left(z^{4}+z^{-22}\right)$, or $105z^7\left(z^{-3}+z^{-29}\right)$. The
next DSP receives $x[4]z^{-2M}$ in cascade input and the same $x[2]z^{-3M}$
generating the $1672\left(z^{-11}+z^{-21}\right)z^{-8}$ term,
which is added to the delayed ($z^{-8}$) output of the first DSP.
This same structure is used for the $x[6]$ and $x[0]$ terms, and the
outputs of those two chains are added together separately. Running the
two separate filters in parallel rather than serially requires an
additional adder, but was shown to reduce power by $\sim0.7\,\text{W}$ in total
for all 8 channels due to the number of registers required to delay
inputs in the serial case.

The additional center tap value ($+2^{14}z^{-16}$) is simply
an upshifted value of the original input, and is added into one of the
systolic filters at the appropriate timepoint.

The same configuration is implemented for each of the 8 samples per
system clock. For the downsample block, since the output is decimated
afterwards, only the even samples are implemented (as a polyphase
filter). For the upsample block, since the decimated odd inputs are
zero-stuffed prior, only the odd samples are implemented and the
even samples (which only have the center tap) are simply delayed
to align to the odd samples.

The frequency response of this halfband filter is shown in
Fig.~\ref{fig:halfbandFreqResp}. The limited rejection near Nyquist
results in some aliasing after decimation for physical signals, but
because the aliasing is equivalent in all channels, experiment simulations
confirmed the overall beamforming is not affected.

\subsection{Matched filter block}

\begin{figure}
\begin{center}
\includegraphics[width=0.9\linewidth]{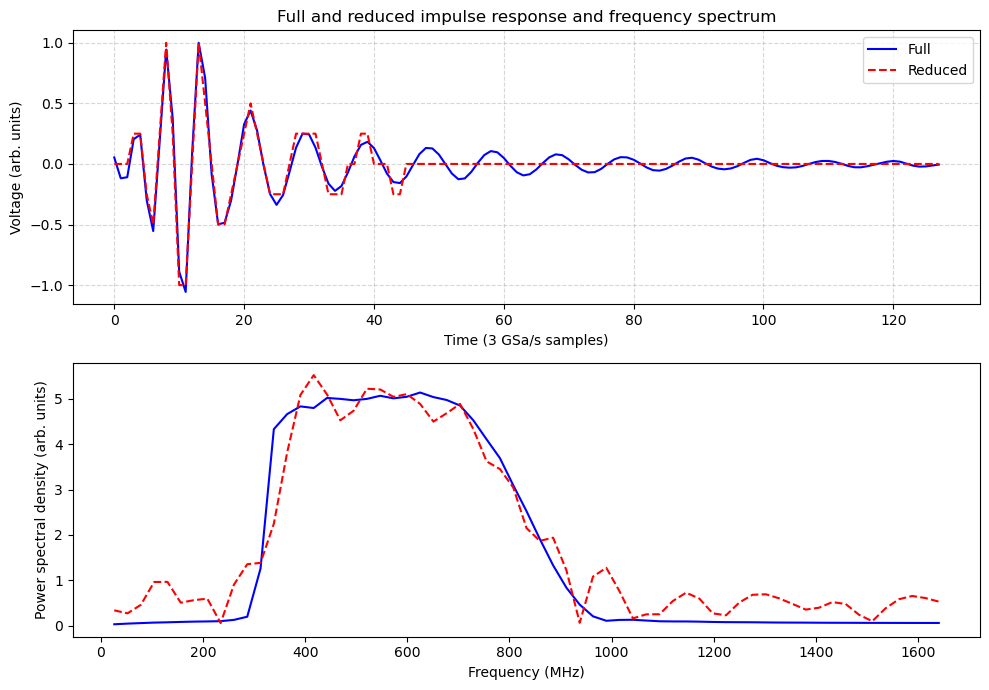}    
\end{center}
\caption{System impulse response (after halfband filter) and reduced version generated by
rounding to the nearest power of 2 for use in the multiplierless
matched filter.}
\label{fig:matchedimpulse}
\end{figure}

The matched filter block compensates for the combined impulse response of the 
antenna and front-end chain, thereby improving the trigger SNR. To minimize
power and resource usage, we approximate this response with a multiplierless
filter with powers-of-2 coefficients. This is an efficient approximation
as the impulse response has an approximately exponential falloff in voltage.
The impulse response was normalized to the peak value, and individual samples were
rounded to $0$, $\pm\frac{1}{4}$, $\pm\frac{1}{2}$, or $\pm1$.
A comparison of the normalized impulse response (post-halfband filter)
and the simplified impulse response in both the time and frequency
domain is shown in Fig.~\ref{fig:matchedimpulse}. The reduced response is
then decimated by a factor of 2 to match the $1500~\text{MSa}/\text{s}$
sample rate (SSR factor $M=4$) in the intermediate part of the processing chain,
time-reversed to create a matched filter, and then scaled down by $2^{5}$
for a gain closer to unity. The remaining gain variation is unimportant
as it is absorbed in the AGC-BR block.

The transfer function of the matched filter at the decimated 
sample rate is given by 

\begin{align*}
H\left(z\right) = \frac{1}{2^{5}} & (-1+z^{-3}-z^{-5}+z^{-7}+z^{-8} \\
                  & -z^{-9}-z^{-10}+z^{-11}+z^{-12}-z^{-13}   \\
                  & -2z^{-14}+2z^{-15}-4z^{-17}+4z^{-18} \\
                  & -2z^{-19}+z^{-20}) \\
\end{align*}

The filter is then simplified by pre-computing $P=\left(1-z^{-1}\right)$
for each sample. The computation of $P$ requires 4 adders at $M=4$, but
reduces the number of terms from $16M$ to $11M$, significantly
reducing the complexity of the filter. The multiplierless nature of the
filter allows it to be implemented efficiently in the FPGA fabric as
sets of ternary adders, which was found to be more power-efficient than a
DSP-based design.

\subsection{RFI rejection block}

Narrowband interferers previously observed in Antarctica, particularly 
MUOS \cite{oetting2011mobile} at $360-380~\textrm{MHz}$ and South Pole radio
transmitters near $450~\textrm{MHz}$, can generate excessive trigger rates
for the antennas in view. Prior balloon flights of the ANITA experiment
\cite{ALLISON201847} used simple tunable analog notches for a rejection
level of $\sim13~\textrm{dB}$ with a $50~\textrm{MHz}$ $3~\textrm{dB}$ bandwidth.
In PUEO, to mitigate this, up to 2 programmable digital biquad filters are 
implemented in the RFI rejection block, operating at an SSR factor $M=4$.
These sections implement the transfer function

\begin{equation}
\label{eq:iirtransfer}
H\left(z\right) = \frac{b_0+b_1z^{-1}+b_0z^{-2}}
{1-2P\alpha z^{-1}+P^{2}z^{-2}}
\end{equation}

This transfer function represents a filter with two complementary poles at
radius $P$ and angle $\theta$ where $\alpha = \cos\theta$ and two
complementary zeros on the unit circle and an overall gain. This structure
allows for the creation of two simple notch filters, or a single higher-order
band rejection filter.

\subsubsection{Biquad design}

In a supersample rate design, the poles of a biquad are challenging
because of the inherent feedback. We split up Eq.~\ref{eq:iirtransfer}
into a zero-only section
$H_z\left(z\right)=b_0+b_1z^{-1}+b_0z^{-2}$ 
and a pole-only section 
$H_p\left(z\right)^{-1}= 1-2P\alpha z^{-1}+P^{2}z^{-2}$. 
$H_{z}$ is applied first since it acts to reduce the overall dynamic range.

The implementation of $H_p\left(z\right)$ uses pipelining along with
clustered look-ahead and incremental computation \cite{parhi2007vlsi}.
This is mathematically equivalent to adding compensating zeros to the
numerator of the transfer function to increase the order of the
denominator. We use the look-back equation from Feinberg
\cite{feinberg116087} to rewrite $H_p\left(z\right)$ as

\begin{equation}
\label{eq:clusteredlookback}
H_{p}\left(z\right)=
\frac{\sum_{i=0}^{\textrm{N}-1} \chebyu{i}{\alpha}P^{i}z^{-i}}
{1+\chebyu{\textrm{N}}{\alpha}P^{\textrm{N}}z^{-\textrm{N}}-\chebyu{\textrm{N}-1}{\alpha}P^{\textrm{N}+1}z^{-\textrm{N}-1}}
\end{equation}

Eq.~\ref{eq:clusteredlookback} allows us to represent $H_{p}$ from 
sequential samples at an increased delay ($z^{-\textrm{N}}$ and
$z^{-\textrm{N}-1}$ rather than $z^{-1}$ and $z^{-2}$). Here 
$U_{i}(\alpha)$ is the Chebyshev polynomial of the second kind.
We choose sample 0 to be calculated with $z^{-(\textrm{M}-1)}$
and $z^{-M}$ ($N=M-1$) and sample 1 to be calculated with $z^{-M}$
and $z^{-(M+1)}$ ($N=M$). The filter instance for PUEO uses $M=4$, however
the filter implementation itself is parameterized for any $M$ and was
originally tested with $M=8$ in a $3000\,\text{MSa/s}$ implementation.

We separate off the numerator of Eq. \ref{eq:clusteredlookback},
defining

\begin{align*}
    f = & \sum_{i=0}^{M-2}\chebyu{i}{\alpha}P^{i}z^{-i}    \\
    g = & \sum_{i=0}^{M-1}\chebyu{i}{\alpha}P^{i}z^{-i}
\end{align*}

As a difference equation, this is a finite impulse response (FIR) filter,
applied to the input after $H_{z}$, only retaining samples 0 and 1
($f\left[0\right]$ and $g\left[1\right]$). The overall difference equation
for $H_{p}\left(z\right)$ for samples 0 and 1, in matrix form, is therefore

\begin{multline*}
    \begin{pmatrix}
        \ssrsample{y}{0} \\
        \ssrsample{y}{1}
    \end{pmatrix}
    =
    P^{M-1}
    \begin{pmatrix}
        -P\chebyu{M-2}{\alpha} & \chebyu{M-1}{\alpha} \\
        -P^{2}\chebyu{M-1}{\alpha} & P\chebyu{M}{\alpha}
    \end{pmatrix}
    \\
    \begin{pmatrix}
        \ssrsample{y}{0}z^{-M} \\
        \ssrsample{y}{1}z^{-M}
    \end{pmatrix}
    +
    \begin{pmatrix}
        \ssrsample{f}{0} \\
        \ssrsample{g}{1}
    \end{pmatrix}
\end{multline*}

This difference equation is then pipelined by substitution, 
creating a final difference equation of

\begin{equation}
\label{eq:finaldifference}
    \begin{pmatrix}
        \ssrsample{y}{0} \\
        \ssrsample{y}{1}
    \end{pmatrix}
    =
    \begin{pmatrix}
        C_0 & C_1 \\
        C_2 & C_3
    \end{pmatrix}
    \begin{pmatrix}
        \ssrsample{y}{0}z^{-2M} \\
        \ssrsample{y}{1}z^{-2M}
    \end{pmatrix}
    +
    \begin{pmatrix}
        F \\
        G
    \end{pmatrix}
\end{equation}

where
\begin{align*}
    F = &
        -P^{M}\chebyu{M-2}{\alpha}z^{-M}\ssrsample{f}{0}+ \\
        & P^{M-1}\chebyu{M-1}{\alpha}z^{-M}\ssrsample{g}{1}+
         \ssrsample{f}{0} \\
    G = &
        -P^{M+1}\chebyu{M-1}{\alpha}z^{-M}\ssrsample{f}{0}+ \\
        & P^{M}\chebyu{M}{\alpha}z^{-M}\ssrsample{g}{1}+
        \ssrsample{g}{1} \\
    C_0 = &
        P^{2M}\left(\left(\chebyu{M-2}{\alpha}\right)^{2} -
                   \left(\chebyu{M-1}{\alpha}\right)^{2}\right) \\
    C_1 = &
        P^{2M-1}\chebyu{M-1}{\alpha}\left(
        \chebyu{M}{\alpha}-\chebyu{M-2}{\alpha}
        \right) \\
    C_2 = &
        P^{2M+1}\chebyu{M-1}{\alpha}\left(
        \chebyu{M-2}{\alpha}-\chebyu{M}{\alpha}
        \right) \\
    C_3 = &
        P^{2M}\left(\left(\chebyu{M}{\alpha}\right)^{2} -
                   \left(\chebyu{M-1}{\alpha}\right)^{2}
             \right) 
\end{align*}

Eq.~\ref{eq:finaldifference} is a pair of coupled infinite impulse
response (IIR) filters where the computation can take up to 
two clock cycles.

\begin{figure}
    \begin{center}
    \includegraphics[width=0.9\linewidth]{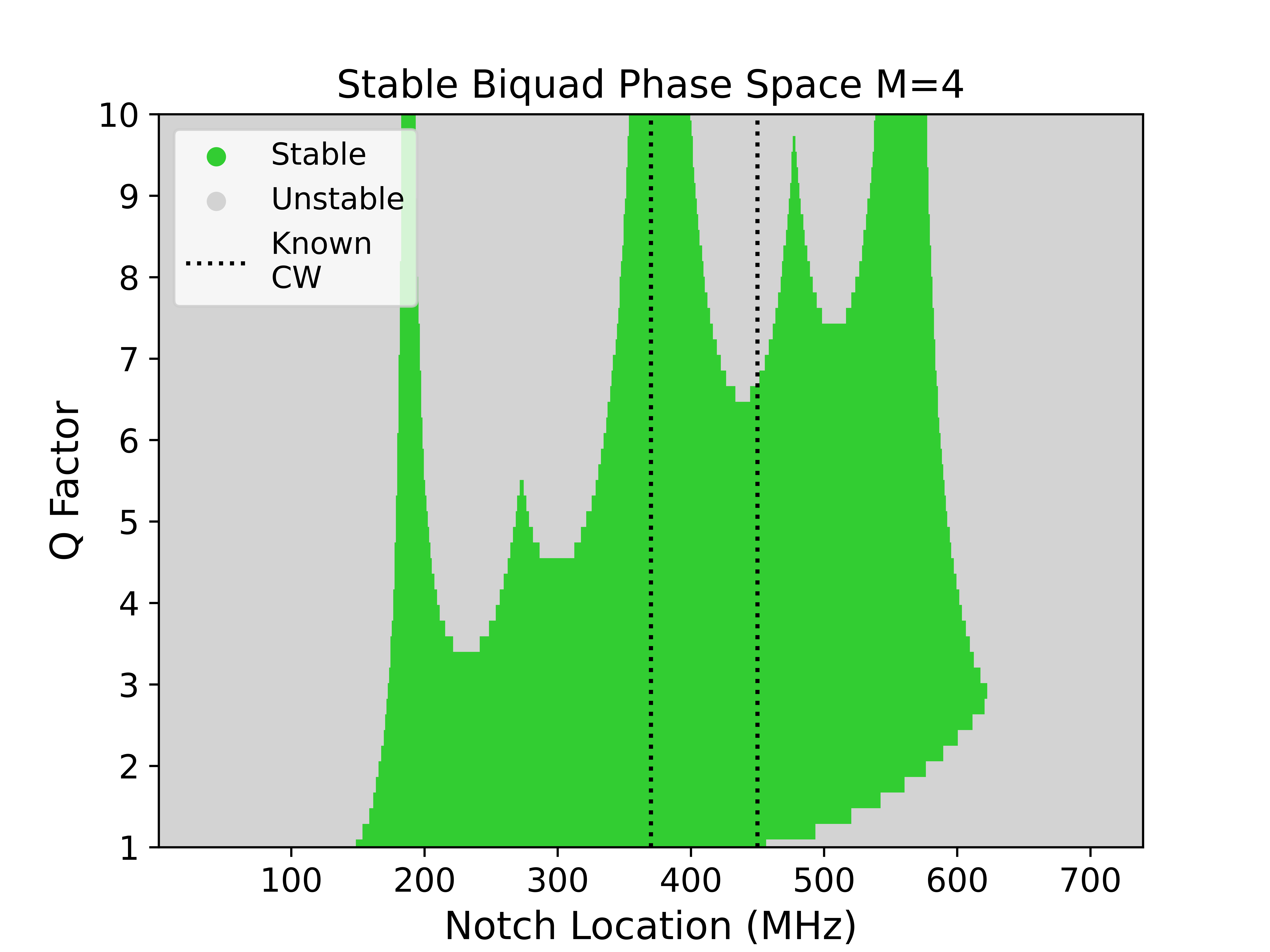}
    \end{center}
    \caption{Stable phase space for a biquad with $M=4$ when used
    as a simple notch with nominal frequencies indicated with a
    vertical line. In the primary PUEO trigger band of 
    $300-750\,\text{MHz}$ the primary interferers occur in the 
    $375-500\,\text{MHz}$ where a good region of stability exists for
    moderate-width notches as used in similar flights.}
    \label{fig:stablebiquad}
\end{figure}

The clustered look-ahead implementation here is not guaranteed
to produce a stable filter, however the regions of stability
do include the primary bands where RF interference was previously
seen. When used as a simple notch, the stable frequency/Q phase
space is shown in Fig.~\ref{fig:stablebiquad}. The main limitations
occur outside of $150\,\text{MHz}<f<600\,\text{MHz}$. To equal the
performance of the analog notches used in ANITA, an effective
$Q$ of $5-7$ would be needed, all of which would be in the stable
region for this design.

\subsubsection{Biquad implementation}

$H_{z}\left(z\right)$, $f$, and $g$ are all straightforward FIR filters,
implemented in a cascade with serial coefficient programming.
The $f\left[n-1\right]$ and $g\left[n-1\right]$ terms in the definition
of $F$ and $G$ are calculated in an additional DSP in the cascade with
the prior DSP's output looped to its input. This term is the critical
path in the $F$ and $G$ calculation, but could be trivially pipelined
by breaking the DSP cascade. Finally, an additional pair of DSPs
calculate the $g$ term in $F$ and the $f$ term in $G$. In total,
the $H_z$ calculation takes $2M$ DSPs, the $f$/$g$ calculation takes $M-2$/$M-1$
DSPs respectively (since the $i=0$ term in $f$/$g$ is a simple addition),
and $F$/$G$ take 2 DSPs each.

\begin{figure}
    \begin{subfigure}{\linewidth}
    \begin{center}
    \includegraphics{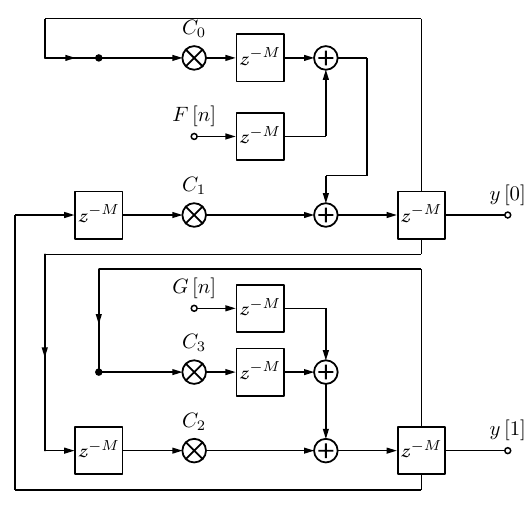}
    \end{center}
    \caption{Biquad IIR structure}
    \label{fig:biquaddesign}
    \end{subfigure}
    \begin{subfigure}{\linewidth}
    \begin{center}
    \includegraphics[width=0.9\linewidth]{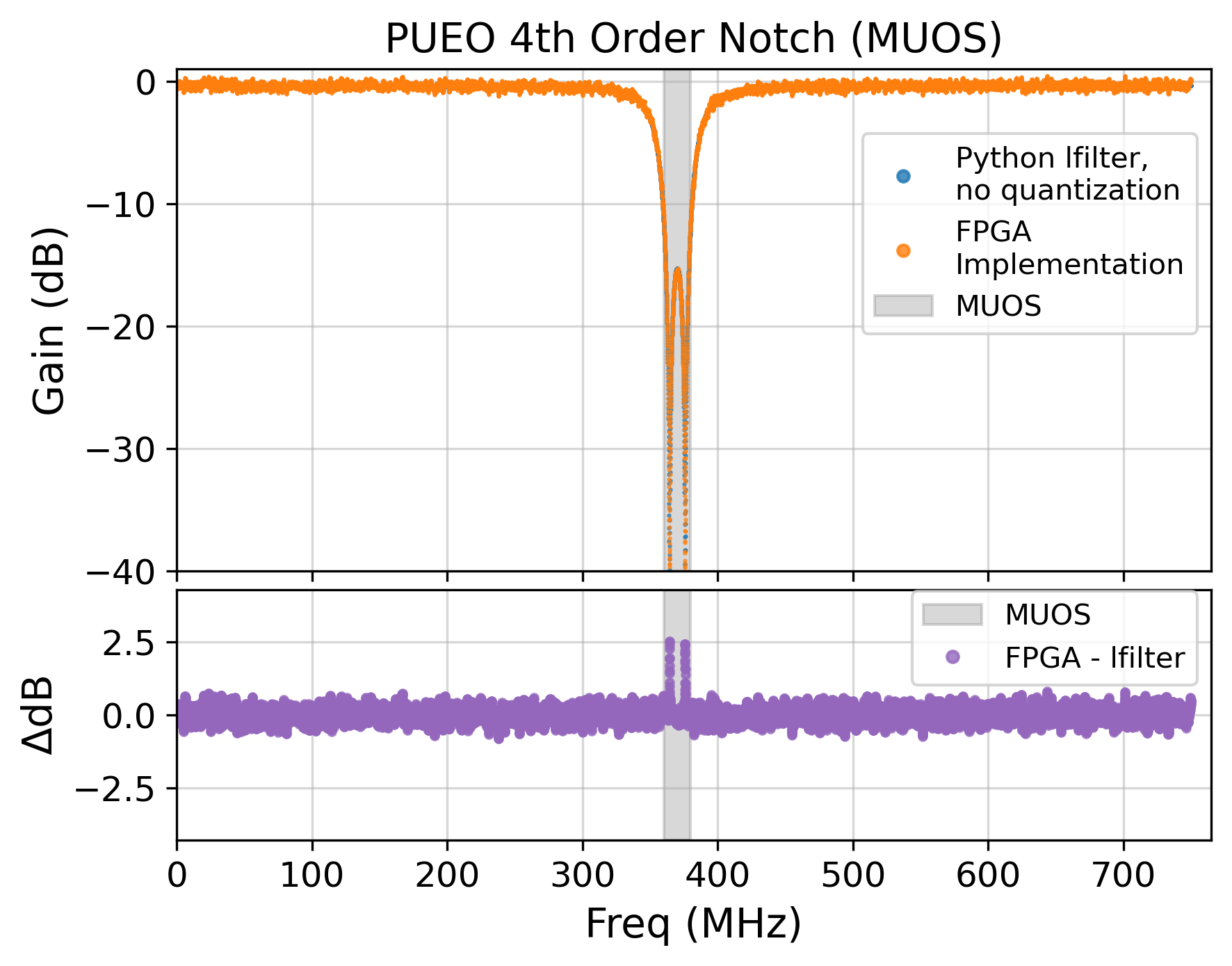}
    \end{center}
    \caption{Example implemented frequency response vs idealized}
    \label{fig:biquadresp}
    \end{subfigure}
    \caption{Biquad IIR structure and example frequency response
    using the implemented design versus floating-point idealization. 
    FPGA implementation frequency response was measured by injecting
    Gaussian noise and examining the output.
    The IIR section contains the critical paths of the filter
    (through the $C_{0}$/$C_{3}$ multiplier) and required specific
    pipeline register choices to meet timing. The example frequency
    response uses 2x biquads to provide a 4th order Chebyshev
    bandstop from $355-385~\textrm{MHz}$.}
\end{figure}

The IIR section consists of a cascade of 4 DSPs with pipeline
registers carefully chosen to meet timing, as shown in 
Fig.~\ref{fig:biquaddesign}. Each sample consists of 2 DSPs, with the
output of the second DSP routed back to the input of the first
with a pipeline register after the multiplier. This feedback is
the critical  fabric routing path of the biquad and
limits this design to a maximum of $\sim425\,\text{MHz}$ with
the given device. The second DSP takes in the other sample's output
with the pipeline register at the input to the DSP. This
maximum speed is independent of the supersample rate factor $M$ -
while the PUEO implementation currently uses $M=4$ (so a $1500\,\text{MSa/s}$
operating frequency) previous versions successfully ran at $M=8$.

The final incremental computation section calculates $y\left[i\right]$
for $i=2\ldots M$ using the original difference equation
$y\left[i\right]=x\left[i\right]-P\cos\theta y\left[i-1\right]-P^{2}y\left[i-2\right]$.
This consists of $2M-4$ DSPs arranged in a cascade to both allow
serial coefficient programming and input sharing between adjacent
samples. The input $x\left[i\right]$ terms needed are propagated through
the $f$ and $g$ calculation sections to allow delays to be shared
between the two portions of the calculation.

An example of the frequency response of the RFI rejection section
programmed as a 4th order bandstop is shown in Fig.~\ref{fig:biquadresp}.
The DSP resource usage for the biquad breaks down as

\begin{itemize}
    \item $H_{z}$ -- $2M$
    \item $F$ and $G$ -- $2M+1$
    \item IIR and incremental computation -- $2M$
\end{itemize}

or $6M+1$ total DSPs (25 for $M=4$). This low DSP count demonstrates the
necessity of the high complexity of the IIR implementation. The best-case
FIR implementation (using a notch at $375~\textrm{MHz}$ only, so half
the coefficients are approximately zero) would require over 48 total DSPs
to reach the $50~\textrm{MHz}$ notch bandwidth.


The RFI rejection section can also be placed into a bypass mode
where the overall transfer function is switched to $H\left(z\right)=1$
by swapping the $b_{1}$ coefficient to 1 and forcing the multiplier
outputs in the DSPs sequentially into reset. The overall trigger
is blocked during these transition periods to prevent glitching
issues. These transitions can be commanded globally to allow the
notches to only apply to specific channels as needed.

\begin{figure}
    \begin{center}
    \includegraphics[width=0.9\linewidth]{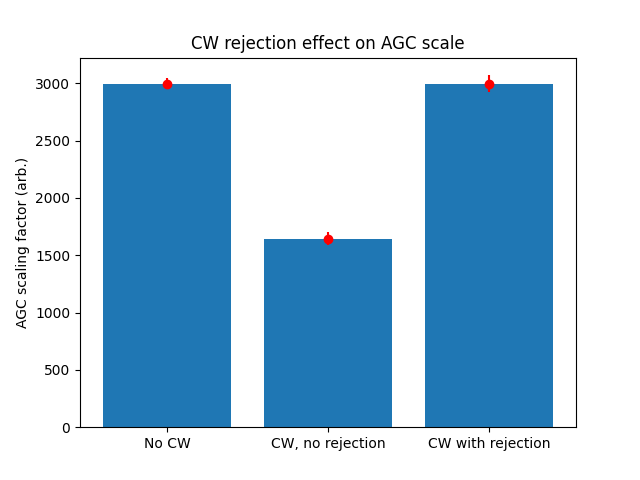}        
    \end{center}
    \caption{\label{fig:cw_mitigation_testing}RFI mitigation testing.
    Injecting a $370~\mathrm{MHz}$ sine wave at $\sim1.75\times$
    thermal RMS with no mitigation resulted in a $\sim1.8\times$
    decrease in the AGC-BR scale to compensate. Enabling the CW
    rejection returned the scaling to its prior value.
    }
\end{figure}

Preflight tests of the RFI rejection were performed by viewing
the scale value from the AGC-BR block (see Sec.~\ref{sec:agc_br}).
A sine wave at $370~\mathrm{MHz}$ was injected into 6 channels
at an amplitude of $\sim1.75\times$ the thermal noise RMS. When
enabled, the AGC-BR scale returned to its no-RFI value,
as shown in Fig.~\ref{fig:cw_mitigation_testing}.

Due to thermal constraints during the 2025-2026 PUEO flight, the system
was operated only with firmware with RFI rejection disabled. In practice,
this primarily affected the subset of phi sectors in which RFI became
significant relative to thermal noise, which amounted to approximately
17\% during this flight. This relatively small fraction was due to the
atypical flight path, which generally kept multiple MUOS satellites out of
view, and to South Pole policy changes that reduced the transmit power of
the $450~\textrm{MHz}$ transmitter when operationally viable.

\section{\label{sec:agc_br}Automatic gain control and bit reduction block (AGC-BR)}

The last block before the beamforming section is the AGC-BR block,
which normalizes each of the channels based on thermal noise
statistics and reduces the overall dynamic range. Prior studies
\cite{xie2021improving} indicated no loss in trigger efficiency and
insensitivity to overall scale at a total dynamic range corresponding
to 5 bits. We therefore reduce to a symmetric signed 5-bit representation 
-- that is, values are represented as a two's complement $x^{\prime}=x-0.5$. 
The scaling of the AGC-BR block targets an output RMS of $4$, meaning 
$1\,\text{LSB}=0.25\sigma$ with zero mean -- as an example, 0 represents all
values from $0\dots0.25\sigma$ (on average $0.125\sigma$) and -1 represents all 
values from $-0.25\dots0\sigma$ (on average $-0.125\sigma$). The symmetric 5-bit
representation means that the output now represents values between $\pm3.875\sigma$.
Values outside this range are clipped to the maximum/minimum values.

In addition to the signed outputs, the AGC-BR block also generates
the absolute value and $x>2\sigma$ and $x<-2\sigma$ (tail fraction)
indicators. Because of the symmetric representation, both of these
outputs are trivial, as
$\text{abs}\left(x\left[4:0\right]\right)=x\left[4\right]\oplus x\left[3:0\right]$,
$x<-2\sigma =x\left[4\right]\cdot \overline{x\left[3\right]}$ and
$x>2\sigma = \overline{x\left[4\right]} \cdot x\left[3\right]$.

The 4-bit absolute value is mapped to its true square and
accumulated using a custom-logic adder and the tail fraction
indicators increment counters over a full AGC-BR measurement period.
Once the measurement period is complete, a per-channel control loop
increments or decrements the scaling factor if the square is outside of a
target range, and increments or decrements an offset based on the
difference of the tail fraction counters. The simple control loop
does result in a slight remaining variation of $\sim1\%$, which is
sufficient given the overall variation in the analog signal path.

\section{Beamforming and trigger}

The output of the signal-processing chain is beamformed into 48 virtual antenna 
beams. For each beam, we compute a simple signal envelope by squaring the samples
and summing them over a short sliding window; threshold crossings then generate
the trigger.

The 48 total beams consist of 35 8-channel beams spanning from 
$\sim0^\circ$ to $\sim-17^\circ$ in zenith and 13 6-channel beams 
centered at $4^{\circ}$ and $-20^{\circ}$ in zenith, both groups
roughly spanning $\pm30^{\circ}$ in azimuth. The 6-channel beams
utilize only the lower antennas, allowing the less signal-rich
zenith extremes to be covered with fewer beams.

\subsection{Beamforming coherent sum}

The 5-bit samples from each of the 8 channels are stored in delay elements
32 samples deep to allow for beamforming. The layout of the
PUEO antennas encourages beams formed from common \emph{sub-beams}
from the 3 bottom ring antennas in each phi sector (termed the
``left'' and ``right'' sub-beams), as they have short baselines
and thus common time delays between multiple beams, due to the 
quantization of time steps. The two antennas in the
top ring are similarly combined (``top'' sub-beam), and then the
left, right, and top sub-beams are delayed relative to each other
and summed to form the final 8-channel collection of 35 total beams.
The 6-channel beams (without the large baseline from the top antennas)
were formed by adding only the left and right sub-beams.

Each of the sub-beams as well as their final combination into full
beams are added using ternary adder structures, since no 
frequency-dependent amplitude or phase correction per antenna is needed due to
the stable beam characteristics of the quad-ridged horn. The unused
addend in the top antenna adder is used to correct the offset in the
symmetric signed representation (and in the 6-antenna beams, the unused
addend in the final ternary adder is used). This coherently
summed signal now ranges over $\pm124$ ($\pm31\sigma$) for the 
8-antenna beams and $\pm93$ ($\pm23.25\sigma$) for the 6-antenna beams.

\subsection{Square and envelope formation}

Once the individual 8-bit samples for each beam are formed,
a simple envelope is generated by squaring each sample
and computing an 8-sample sum every 4 samples.

The squarer is based on the signed 8-bit logic given by 
Wires et al. \cite{wires831900}, but optimized for the
fabric of the UltraScale family and their generic 5-input,
2-output lookup table (LUT).

The squarer adder tree is structured overall as a ternary adder,
with a 3:2 compressor embedded into the LUT, with one (sum) output
feeding into the carry chain and the second (carry) output
routed to the next bit. Additionally, two 4-bit secondary terms
were computed from the sum of several partial products containing
only 5 common inputs, which allowed them to be directly computed
without a carry chain.

Several bits in the ternary adder were left with unused outputs,
allowing other secondary terms to be generated,
reducing the total size to a 10-bit ternary adder (10 LUTs) and
4 additional auxiliary LUTs for the secondary terms (14 LUT6s total), 
fitting in less than 2 UltraScale slices. For comparison, a
synthesis-generated 8-bit square generated over 8 slices of logic
($>64\,\mathrm{LUT6}$), a logic reduction of over a factor of 4.
The worst critical path of the logic consists of 4 levels of logic
(3 LUT propagation times plus the carry chain), corresponding to
$\sim0.75\,\text{ns}$ in the device used, significantly less than
the synthesis-generated design.

This overall method is similar to that described independently by Bui et al.
\cite{bui2014additional} except for signed cases, but optimized
to ensure as much logic could be integrated into the adder LUTs
as possible by abandoning generality. While other optimized
squarers for FPGAs have been presented \cite{bottcher2022resource},
these were focused on building efficient larger-width squarers 
or for direct ASIC implementation, and the detailed size of small 
squarers was not clear. The aggressive optimization here is most
appropriate because of the large number of squares ($384~\mathrm{total}$) 
required -- the optimized square reduced the slice count of the
beamforming trigger by $56\%$ alone.

\begin{figure}
\resizebox{0.95\linewidth}{!}{
$\begin{array}{cccccccccc}
s_{14} & s_{13} & s_{12} & s_{11} & s_{10} & s_{9} & s_{8} & s_{7} & s_{6} & s_{5}\\
\hline 
T_{14} & T_{13} & \overline{a_{7}a_{4}} & T_{11} & \overline{P_{72}} & \overline{a_{7}a_{1}} & \overline{a_{7}a_{0}} & a_{6}a_{0} & a_{5}a_{0} & T_{5} \\
 &  & c_{3} &  & c_{1} & c_{0} & a_{6}a_{1} & d_{1} & d_{0} & P_{40}\\
 &  & \overline{a_{7}a_{3}}c_{2} &  & \bar{d}_{3}a_{4} & d_{3} & d_{2}
\end{array}$
}

\[
\begin{array}{cccc}
c_{3} & c_{2} & c_{1} & c_{0}\\
\hline a_{6}\bar{a}_{5} & a_{6}a_{4} & a_{6}a_{3} & a_{6}a_{2}\\
 & a_{5}a_{4} & a_{5}\bar{a}_{4} & a_{5}a_{3}\\
 &  &  & a_{5}a_{2}\left(a_{3}\lor\bar{a}_{4}\right)
\end{array}
\]

\[
\begin{array}{cccc}
d_{3} & d_{2} & d_{1} & d_{0}\\
\hline a_{4} & a_{5}a_{2}\oplus\overline{a_{4}\bar{a}_{3}} & a_{5}a_{1} & a_{3}\bar{a}_{2}\\
 &  & a_{4}a_{2} & a_{4}a_{1}\\
 &  & a_{3}a_{2} & a_{3}a_{2}a_{1}
\end{array}
\]

\begin{align*}
T_{5} = & a_{3}a_{1}\oplus a_{3}a_{2}a_{0}\oplus a_{2}a_{1} \,(\text{internal logic to bit 5})  \\
T_{11} = & \overline{a_{7}a_{3}}\oplus c_{2}\,(\text{internal logic to bit 11}) \\
T_{13} = & \overline{a_{7}a_{5}}\oplus a_{6}a_{5}\,(\text{internal logic to bit 13}) \\
T_{14} = & \overline{a_{7}\bar{a}_{6}}\oplus\bar{a}_{7}a_{6}a_{5}\,(\text{internal logic to bit 14}) \\
P_{40} = & a_{4}a_{0}\,(\text{secondary output from bit 7}) \\
P_{72} = & a_{7}a_{2}\,(\text{secondary output from bit 11}) \\
s_{4} = & a_{2}\overline{a_{1}} \oplus a_{3}a_{0} \oplus a_{2}a_{1}a_{0}\,
(\text{secondary output from bit 5}) \\
s_{3} = & a_{2}a_{1} \oplus a_{1}a_{0}\,(\text{secondary output from bit 6}) \\
s_{2} = & a_{1}\overline{a_{0}}\,(\text{secondary output from bit 14}) \\
s_{1} = & 0 \\
s_{0} = & a_{0} \\
\end{align*}

\caption{Optimized two's complement 8-bit square logic for 5-bit
input, 2-bit output LUTs, with the inputs represented as $a_{7:0}$
and the outputs as $s_{14:0}$. The overall logic consists of a 10-bit
ternary adder structure (consisting of the terms under $s_{14:5}$) and 
several pre-computed partial products and secondary terms. The two 4-bit
secondary terms ($c_{3:0}$ and $d_{3:0}$) are logically outputs of
ternary adders, but since the terms are all derived from only 5 inputs
these are implemented as direct LUTs. Note that $\lor$ indicates a logical
OR operation and $\oplus$ indicates a logical XOR.}
\end{figure}

After the square of each sample was generated, the final step
in the envelope trigger was to compute an 8-sample sum every 4 samples,
which corresponds to 2 samples every system clock. Because the sums
are thresholded, only the larger of the two matters: therefore,
we compute the sum of samples 4 to 7,  compare to the previous
clock cycle, and retain only the larger of the two. This max sum
is then added to the sum of samples 0 to 3, generating the max of
$\sum_{i=4}^{7}x\left[i\right]z^{-8}+\sum_{i=0}^{3}x\left[i\right]$
and $\sum_{i=0}^{7}x\left[i\right]$. This value is then thresholded
twice -- one threshold is used to generate the trigger output of
that beam, and the second threshold is used as a subtrigger
monitoring threshold to allow observation of the beam behavior with
higher statistics.

Thresholding is done with Xilinx DSPs configured in a 2x24-bit SIMD
configuration, handling 2 beams at a time. The entire beamforming
section is implemented in a DSP cascade to allow the thresholds to
be updated using the coefficient cascade feature of the DSPs.

\subsection{L1 Trigger}

The final (L1) trigger provided to the Trigger Unit for RF (TURF)
global PUEO trigger system is generated from the OR of all 48 individual
thresholded beams and quantized to a 3-clock (24 total samples)
interval. The trigger is provided as a 12-bit trigger timestamp, 
corresponding to a $98,304~\textrm{sample}$ ($32.768\,\mu\textrm{s}$) 
readout signal buffer for absorbing the trigger formation latency.

An additional 8-bit metadata is also transmitted with each trigger. 
The 48 total beams are placed in one or more of 8 overall groups
corresponding to their direction. If a beam in one of those groups 
passes the threshold, its corresponding bit in the metadata is set.

\subsection{L2 and Global Trigger}

At the TURF, when an L1 trigger is received, the trigger and metadata
for each SURF at that timestamp are marked valid in a rolling
$4096~\text{sample}$ buffer. The values from this buffer are read
out a programmable time latency later, which allows the L1 triggers
to be received at different times from each SURF.

If two bits in adjacent SURF metadata corresponding to the same
beam group direction are seen within $16~\textrm{ns}$, this 
generates an L2 trigger, which physically corresponds to a 
signal in a $30^\circ$ section of the payload roughly centered 
between the adjacent SURF antenna groups. Each
individual L2 trigger could be masked off to eliminate noisy
directions.

An unmasked L2 trigger then generates a global system trigger.
The TURF returns the readout timestamp to all SURFs, and a
$1024~\text{sample}$ buffer is read out from the signal buffer
and transmitted to the TURF and then to the science flight computer.

The individual beam trigger rates during the 2025-2026 flight
were $\sim650\,\mathrm{Hz}$ during normal operation (tuned to
produce the desired total experiment rate), which resulted in a
typical total L1 rate of $\sim28\,\mathrm{kHz}$ per SURF,
and an L2 rate per $30^\circ$ section of $\sim4-5~\mathrm{Hz}$
for a total experiment trigger rate of $\sim 100~\mathrm{Hz}$.

\section{Performance}

\begin{table}
\begin{center}
\begin{tabular}{ |c|c|c| } 
 \hline
    & Slices & DSPs \\
    \hline
    Total available on device & 53160 & 4272  \\
    Signal processing chain (excluding biquads) & 2767 (5.2\%) & 584 (13.7\%) \\
    Biquad (each, up to 2) & 800 (1.5\%) & 200 (4.7\%) \\
    Beamforming trigger (48 total beams) & 2120 (4.0\%) & 120 (2.8\%) \\
 \hline
\end{tabular}
\end{center}
\caption{
\label{tab:resource} Resource usage for the trigger chain.
Each slice consists of 8 LUTs and 16 registers. The beamforming
trigger section also consists of resources to monitor the trigger
rates of each beam for threshold monitoring.
}
\end{table}

\subsection{Resource and Power Usage}
The overall major resource usage breakdown for the trigger system
at the SURFs is shown in Table~\ref{tab:resource}. In total, with 2 biquads,
the trigger system uses $\sim12\%$ of the CLBs in the device and $\sim26\%$
of the DSPs. Note that the resource usage of the L2 trigger at the TURF is
negligible.

The SURF firmware was designed to allow the trigger portion to be activated 
after initialization, which enables a direct power measurement of 
the active portion of the trigger. At
$35^{\circ}~\mathrm{C}$, the trigger system, excluding the biquads,
used $\sim 5\,\mathrm{W}$ of power, and the biquads additionally
added $\sim 1\,\mathrm{W}$ each in bypass, and an additional
$\sim1\,\mathrm{W}$ when active. 

Synthesis and implementation times for the SURF firmware were additionally
optimized by relaxing as many timing paths in the system clock domain as
possible, isolating the trigger portion of the firmware from the
processing system (freeing up placement) and grouping the squaring logic
in the same device slice. These optimizations kept build times on the order of
one hour, allowing for rapid development cycles.

\subsection{Trigger Efficiency}

PUEO's trigger behavior can be tested in full operation using simulated Askaryan RF signals, 
produced using a custom calibration instrument built around a Xilinx ZCU216 RFSoC evaluation board (ZCU216) by the PUEO collaboration. 
This calibration instrument, termed the ``RF box'', can produce up to 16 such signals with 
relative delays between channels that match the arrival times of a plane wave incident on 
PUEO's antennas from a given direction. The RF box includes variable attenuators that can be 
used to scale the signal amplitude. The RF box is attached in place of the antennas 
for two adjacent SURFs, feeding signals first into a 30 dB attenuator and then into PUEO's full RF chain 
and into the DAQ. The RF box is set to simulate a signal arriving from a direction visible to the 
beams of both of the SURFs under test. Details on the RF box will be presented elsewhere.

\begin{figure}
    \begin{center}
    \includegraphics[width=0.9\linewidth]{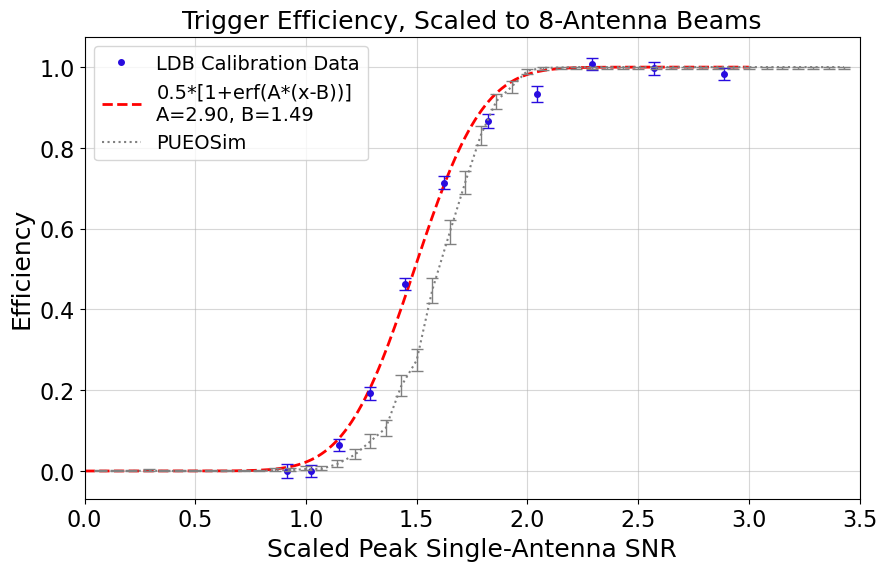}        
    \end{center}
    \caption{\label{fig:efficiency_beamformer} Beamform trigger efficiency, calculated from data taken in the launch facility
    hangar using the RF box. Single-antenna SNR has been scaled by $\frac{3}{4}$ to represent an 8-antenna beam, as described in text. PUEO's Monte Carlo instrument simulation, PUEOSim, is plotted in gray for comparison.
    }
\end{figure}

Measurements of the trigger efficiency were made using this test configuration while on the 
ground at McMurdo Station's Long Duration Balloon (LDB) launch facility, prior to the instrument flight. 
Due to space constraints in the launch facility hangar, only the 12 lower-ring antenna paths were attached to the RF box.
The top-ring front-end amplifiers were terminated, providing a similar noise environment without the
simulated signal. The signal direction was chosen to be arriving from $-15^{\circ}$ in elevation and $-5^{\circ}$ in azimuth, 
relative to the midline between the four instrumented phi sectors. Due to time constraints, this was the only 
exhaustively measured direction during this test, but was chosen to be representative of coverage in the most likely
arrival directions. Prior to data taking, the flight software threshold control servo was allowed to adjust the 
trigger threshold to a level that provided flight-like trigger rates. Once this threshold value settled, the RF box was set to send signals at 10 Hz, significantly below the servo's target rate.

Various signal sizes were measured in 11 approximately 2 minute-long flight-like data taking runs by changing the programmable attenuation
of the RF box between runs. The single-antenna SNR for the lowest attenuation, where the signals greatly exceeded the noise level, was
calculated using the definition

\begin{equation}
SNR = \frac{ADC_{\textrm{max}} - ADC_{\textrm{min}}}{2 \times ADC_{\textrm{RMS}}}    
\end{equation}

As the attenuation was increased, the single-antenna SNR was calculated by scaling this initial measurement appropriately. In beamforming, noise adds as $\sqrt{N}$, where $N$ is the number of channels being summed, whereas signals add as $N$. Therefore, since only 6 antennas were instrumented per SURF but all 8 had equivalent noise environments, the equivalent single-antenna SNR for an 8-antenna beam can be calculated from this measurement by scaling the SNR by $\sfrac{\frac{6}{\sqrt{8}}}{\frac{8}{\sqrt{8}}} = \frac{3}{4}$.

Each event in the data had its time past the GPS second calculated, termed its ``subsecond''. When these subseconds are binned, 10 spikes corresponding to the RF box's 10 Hz pulsing rate are clearly visible for the highest signal amplitude case. The subsecond bins in which these spikes occur can be summed to produce the total number of measured triggers during the expected arrival of the calibration signals (plus noise). Relative clock drift between the two systems was found to be negligible for the length of the test, so these timing windows are reused for low-amplitude signals as well. Noise is subtracted from this sum by measuring the number of events gathered in an equal-width bin offset from the expected arrival subsecond times. The efficiency is then calculated by dividing the number of remaining events from the total number of expected events for the length of the run. This result is shown in Fig.~\ref{fig:efficiency_beamformer}. Note that the efficiency as calculated here is not a measurement of PUEO's trigger efficiency during flight, which will be covered elsewhere.

\section{Summary}

We have presented a fully digital, low-power trigger system for PUEO that combines
matched filtering, dynamic normalization, RFI mitigation, and real-time beamforming
in the RFSoC fabric. In laboratory tests, the system achieved a 50\% trigger efficiency
at a scaled single-antenna SNR of approximately 1.5 while consuming $5-9~\textrm{W}$ in
the trigger portion and using about 12\% of slice logic and 26\% of DSP resources,
depending on configuration.

\section{Acknowledgments}

The PUEO Collaboration acknowledges the significant
contributions to this manuscript by Patrick S. Allison
and Lucas W. Beaufore.

\input{acknowledgements.tex}


\bibliographystyle{IEEEtran}
\bibliography{refs}

\end{document}

%% file: pueoieee_authors.tex

\author{
Q.~Abarr\orcidlink{0000-0002-9063-2890},
J.~Alfaro,
P.~Allison\orcidlink{0000-0001-8141-2653},
J.~Alvarez-Mu\~niz\orcidlink{0000-0002-2367-0803},
T.~Anderson,
H.~Barnett,
A.~Basharina-Freshville\orcidlink{0000-0001-5627-7832},
J.~J.~Beatty\orcidlink{0000-0003-0481-4952},
L.~Beaufore\orcidlink{0000-0003-2949-1555},
D.~Z.~Besson\orcidlink{0000-0001-6733-963X},
M.~Betts,
R.~Bose,
D.~Braun,
B.~Chamanbahar,
P.~Chen\orcidlink{0000-0001-5251-7210},
Y.~Chen\orcidlink{0000-0002-8967-4911},
J.~M.~Clem\orcidlink{0000-0001-5681-6883},
T.~Coakley\orcidlink{0009-0005-6862-1860},
A.~Connolly\orcidlink{0000-0003-0049-5448},
K.~Couberly,
L.~Cremonesi\orcidlink{0000-0003-0711-1056},
A.~Cummings\orcidlink{0000-0002-9012-7851},
P.~Dasgupta\orcidlink{0000-0002-9559-4803},
C.~Deaconu\orcidlink{0000-0002-4953-6397},
J.~Flaherty\orcidlink{0000-0003-3609-7361},
C.~Godden,
P.~W.~Gorham\orcidlink{0000-0002-0923-9363},
C.~Hornhuber,
J.~Hoffman,
K.~Hughes\orcidlink{0000-0002-4551-9581},
A.~Hynous,
A.~Jung\orcidlink{0000-0002-0880-4148},
A.~M.~Kofman\orcidlink{0000-0001-5374-1767},
Y.~Ku,
D.~Kullgren\orcidlink{0000-0002-0774-1325},
C.-Y.~Kuo,
P.~Linton\orcidlink{0009-0005-6821-3306},
L.~Lisalda\orcidlink{0000-0002-5202-1642},
T.~C.~Liu,
W.~Luszczak\orcidlink{0000-0003-3085-0674},
S.~C.~Mackey\orcidlink{0000-0002-1414-7236},
Z.~Martin\orcidlink{0009-0003-2075-8644},
K.~McBride\orcidlink{0000-0003-2195-4324},
C.~Miki\orcidlink{0000-0002-8968-7835},
J.~Nam,
R.~J.~Nichol\orcidlink{0000-0003-0557-0443},
A.~Novikov\orcidlink{0000-0002-1086-7252},
A.~Nozdrina,
E.~Oberla\orcidlink{0000-0001-8344-7999},
S.~Prohira\orcidlink{0000-0002-8814-6607},
H.~Pumphrey,
D.~Radomski,
B.~F.~Rauch\orcidlink{0000-0002-1452-4142},
R.~Scrandis\orcidlink{0009-0003-1827-6128},
D.~Seckel,
M.~F.~H.~Seikh,
J.~Shiao,
G.~Simburger,
J.~Tutt\orcidlink{0000-0002-1613-0796},
A.~G.~Vieregg\orcidlink{0000-0002-4528-9886},
S.-H.~Wang\orcidlink{0000-0002-0060-7975},
D.~Washington,
C.~Welling\orcidlink{0000-0003-4531-8058},
P.~Windischhofer\orcidlink{0000-0001-5038-1399},
S.~A.~Wissel\orcidlink{0000-0003-0569-6978},
C.~Xie,
J.~Yao,
R.~Young,
E.~Zas\orcidlink{0000-0002-4430-8117},
A.~Zeolla}

%% file: pueoieee_institutes.tex
\thanks{Q. Abarr, P. Allison, J. J. Beatty, L. Beaufore, T. Coakley, A. Connolly, P. Dasgupta, J. Flaherty, K. Hughes, P. Linton, W. Luszczak and J. Yao are with Ohio State University, Dept. of Physics, Center for Cosmology and AstroParticle Physics, Columbus, OH 43210. (e-mail: allison.122@osu.edu)}
\thanks{J. Alvarez-Mu\~niz and E. Zas are with Universidade de Santiago de Compostela, Instituto Galego de F\'isica de Altas Enerx\'ias (IGFAE), 15782 Santiago de Compostela, Spain.}
\thanks{J. Alfaro, T. Anderson, M. Betts, A. Cummings, D. Kullgren, Y. Ku, J. Tutt, D. Washington, C. Welling, S. A. Wissel and A. Zeolla are with Pennsylvania State University, Physics Dept., Inst. for Gravitation and the Cosmos, University Park, PA 16802.}
\thanks{H. Barnett, A. Basharina-Freshville, C. Godden, R. J. Nichol, H. Pumphrey and C. Xie are with University College London, Dept. of Physics and Astronomy, London, United Kingdom.}
\thanks{D. Z. Besson, K. Couberly, C. Hornhuber, A. Nozdrina, S. Prohira, M. F. H. Seikh and R. Young are with University of Kansas, Dept. of Physics and Astronomy, Lawrence, KS 66045.}
\thanks{R. Bose, D. Braun, J. Hoffman, L. Lisalda, D. Radomski, B. F. Rauch and G. Simburger are with Washington University in St. Louis, Dept. of Physics, McDonnell Center for the Space Sciences, St. Louis, MO 63130.}
\thanks{B. Chamanbahar, C. Deaconu, A. M. Kofman, S. C. Mackey, Z. Martin, K. McBride, E. Oberla, R. Scrandis, A. G. Vieregg and P. Windischhofer are with University of Chicago, Dept. of Astronomy and Astrophysics, Dept. of Physics, Enrico Fermi Inst., Kavli Inst. for Cosmological Physics, Chicago, IL 60637.}
\thanks{P. Chen, Y. Chen, C.-Y. Kuo, J. Nam, J. Shiao and S.-H. Wang are with National Taiwan University, Dept. of Physics, Grad. Inst. of Astrophys., Leung Center for Cosmology and Particle Astrophysics, Taipei, Taiwan.}
\thanks{J. M. Clem, A. Novikov and D. Seckel are with University of Delaware, Dept. of Physics, Newark, DE 19716.}
\thanks{L. Cremonesi is with Queen Mary University of London, School. of Physics and Astronomy, London, United Kingdom.}
\thanks{P. W. Gorham, A. Jung and C. Miki are with University of Hawaii-Manoa, Dept. of Physics and Astronomy, Honolulu, HI 96822.}
\thanks{A. Hynous is with NASA Wallops Flight Facility, Wallops Island, VA, 23337}
\thanks{T. C. Liu is with National Pingtung University, Dept. of Applied Physics, Pingtung City, Taiwan.}

%% file: acknowledgements.tex
PUEO is grateful to be selected as a NASA Pioneers mission funded by NASA grants \#80NSSC20K0775 and \#80NSSC20K0925.
We would like to particularly thank the staff at the Columbia Scientific Balloon Facility, the National Science Foundation, and the Antarctic Support Contractor, whose talented and tireless efforts made PUEO possible.
UChicago additionally thanks the Kavli Institute for Cosmological Physics and the University of Chicago Research Computing Center.
R. Scrandis is supported by NASA FINESST award 80NSSC25K7301.
UCL is supported by the UK's Science and Technology Facilities Council.
D. Besson and D. Seckel thank the NSF for support through the IceCube EPSCoR Initiative (Award ID No. 2019597).
D. Kullgren is supported by an NSF GRFP award DGE1255832.
The Ohio State University thanks the Ohio Supercomputer Center for computational resources.
PUEO data will eventually be made publicly available through the High Energy Astrophysics Science Archive Research Center (HEASARC), a service of the Astrophysics Science Division at NASA/GSFC.

%% file: refs.bib
@IEEEtranBSTCTL{IEEEexample:BSTcontrol,
CTLuse_forced_etal       = "yes",
CTLmax_names_forced_etal = "4",
CTLnames_show_etal       = "3" }

@article{PhysRevLett.99.171101,
  title = {Observations of the {Askaryan} Effect in Ice},
  author = {Gorham, P. W. and Barwick, S. W. and Beatty, J. J. and Besson, D. Z. and Binns, W. R. and Chen, C. and Chen, P. and Clem, J. M. and Connolly, A. and Dowkontt, P. F. and DuVernois, M. A. and Field, R. C. and Goldstein, D. and Goodhue, A. and Hast, C. and Hebert, C. L. and Hoover, S. and Israel, M. H. and Kowalski, J. and Learned, J. G. and Liewer, K. M. and Link, J. T. and Lusczek, E. and Matsuno, S. and Mercurio, B. and Miki, C. and Mio\ifmmode \check{c}\else \v{c}\fi{}inovi\ifmmode \acute{c}\else \'{c}\fi{}, P. and Nam, J. and Naudet, C. J. and Ng, J. and Nichol, R. and Palladino, K. and Reil, K. and Romero-Wolf, A. and Rosen, M. and Ruckman, L. and Saltzberg, D. and Seckel, D. and Varner, G. S. and Walz, D. and Wu, F.},
  collaboration = {ANITA Collaboration},
  journal = {Phys. Rev. Lett.},
  volume = {99},
  issue = {17},
  pages = {171101},
  numpages = {5},
  year = {2007},
  month = {Oct},
  publisher = {American Physical Society},
  doi = {10.1103/PhysRevLett.99.171101},
  url = {https://link.aps.org/doi/10.1103/PhysRevLett.99.171101}
}

@article{PhysRevLett.116.141103,
  title = {Accelerator Measurements of Magnetically Induced Radio Emission from Particle Cascades with Applications to Cosmic-Ray Air Showers},
  author = {Belov, K. and Mulrey, K. and Romero-Wolf, A. and Wissel, S. A. and Zilles, A. and Bechtol, K. and Borch, K. and Chen, P. and Clem, J. and Gorham, P. W. and Hast, C. and Huege, T. and Hyneman, R. and Jobe, K. and Kuwatani, K. and Lam, J. and Liu, T. C. and Nam, J. and Naudet, C. and Nichol, R. J. and Rauch, B. F. and Rotter, B. and Saltzberg, D. and Schoorlemmer, H. and Seckel, D. and Strutt, B. and Vieregg, A. G. and Williams, C.},
  collaboration = {T-510 Collaboration},
  journal = {Phys. Rev. Lett.},
  volume = {116},
  issue = {14},
  pages = {141103},
  numpages = {6},
  year = {2016},
  month = {Apr},
  publisher = {American Physical Society},
  doi = {10.1103/PhysRevLett.116.141103},
  url = {https://link.aps.org/doi/10.1103/PhysRevLett.116.141103}
}

@book{parhi2007vlsi,
  title={{VLSI} Digital Signal Processing Systems: Design and Implementation},
  author={Parhi, Keshab K},
  year={2007},
  publisher={John Wiley \& Sons}
}

@article{xie2021improving,
  title={Improving Radio Frequency Detectors using High Performance Programmable Logic},
  author={Xie, Cheng},
  journal={PoS ICRC2021},
  volume={1028},
  year={2021}
}

@INPROCEEDINGS{feinberg116087,
  author={Feinberg, R.B.},
  booktitle={International Conference on Acoustics, Speech, and Signal Processing}, 
  title={Vectorizing {IIR} Filters}, 
  year={1990},
  volume={},
  number={},
  pages={1045-1048 vol.2},
  doi={10.1109/ICASSP.1990.116087}}

@article{Abarr_2021,
doi = {10.1088/1748-0221/16/08/P08035},
url = {https://doi.org/10.1088/1748-0221/16/08/P08035},
year = {2021},
month = {August},
publisher = {IOP Publishing},
volume = {16},
number = {08},
pages = {P08035},
author = {Abarr, Q. and Allison, P. and Ammerman Yebra, J. and Alvarez-Muñiz, J. and Beatty, J.J. and Besson, D.Z. and Chen, P. and Chen, Y. and Xie, C. and Clem, J.M. and Connolly, A. and Cremonesi, L. and Deaconu, C. and Flaherty, J. and Frikken, D. and Gorham, P.W. and Hast, C. and Hornhuber, C. and Huang, J.J. and Hughes, K. and Hynous, A. and Ku, Y. and Kuo, C.-Y. and Liu, T.C. and Martin, Z. and Miki, C. and Nam, J. and Nichol, R.J. and Nishimura, K. and Novikov, A. and Nozdrina, A. and Oberla, E. and Prohira, S. and Prechelt, R. and Rauch, B.F. and Roberts, J.M. and Romero-Wolf, A. and Russell, J.W. and Seckel, D. and Shiao, J. and Smith, D. and Southall, D. and Varner, G.S. and Vieregg, A.G. and Wang, S.-H. and Wang, Y.-H. and Wissel, S.A. and Young, R. and Zas, E. and Zeolla, A.},
collaboration = {PUEO},
title = {The {Payload for Ultrahigh Energy Observations (PUEO)}: a white paper},
journal = {Journal of Instrumentation}
}

@article{oetting2011mobile,
  title={The {Mobile User Objective System}},
  author={Oetting, John D and Jen, Tao},
  journal={Johns Hopkins APL technical digest},
  volume={30},
  number={2},
  pages={103--112},
  year={2011},
  publisher={JHU/APL}
}

@article{ALLISON201847,
title = {Dynamic tunable notch filters for the {Antarctic Impulsive Transient Antenna} ({ANITA})},
journal = {Nuclear Instruments and Methods in Physics Research Section A: Accelerators, Spectrometers, Detectors and Associated Equipment},
volume = {894},
pages = {47-56},
year = {2018},
issn = {0168-9002},
doi = {https://doi.org/10.1016/j.nima.2018.03.059},
url = {https://www.sciencedirect.com/science/article/pii/S016890021830411X},
author = {P. Allison and O. Banerjee and J.J. Beatty and A. Connolly and C. Deaconu and J. Gordon and P.W. Gorham and M. Kovacevich and C. Miki and E. Oberla and J. Roberts and B. Rotter and S. Stafford and K. Tatem and L. Batten and K. Belov and D.Z. Besson and W.R. Binns and V. Bugaev and P. Cao and C. Chen and P. Chen and Y. Chen and J.M. Clem and L. Cremonesi and B. Dailey and P.F. Dowkontt and S. Hsu and J. Huang and R. Hupe and M.H. Israel and J. Kowalski and J. Lam and J.G. Learned and K.M. Liewer and T.C. Liu and A.B. Ludwig and S. Matsuno and K. Mulrey and J. Nam and R.J. Nichol and A. Novikov and S. Prohira and B.F. Rauch and J. Ripa and A. Romero-Wolf and J. Russell and D. Saltzberg and D. Seckel and J. Shiao and J. Stockham and M. Stockham and B. Strutt and G.S. Varner and A.G. Vieregg and S. Wang and S.A. Wissel and F. Wu and R. Young},
collaboration = {ANITA}
}

@INPROCEEDINGS{wires831900,
  author={Wires, K.E. and Schulte, M.J. and Marquette, L.P. and Balzola, P.I.},
  booktitle={Conference Record of the Thirty-Third Asilomar Conference on Signals, Systems, and Computers (Cat. No.CH37020)}, 
  title={Combined Unsigned and Two's Complement Squarers}, 
  year={1999},
  volume={2},
  number={},
  pages={1215-1219 vol.2},
  doi={10.1109/ACSSC.1999.831900}}

@inproceedings{bottcher2022resource,
  title={Resource Optimal Squarers for {FPGAs}},
  author={B{\"o}ttcher, Andreas and Kumm, Martin and De Dinechin, Florent},
  booktitle={2022 32nd International Conference on Field-Programmable Logic and Applications (FPL)},
  pages={40--46},
  year={2022},
  organization={IEEE}
}

@inproceedings{bui2014additional,
  title={Additional Optimizations for Parallel Squarer Units},
  author={Bui, Son and Stine, James E},
  booktitle={2014 IEEE International Symposium on Circuits and Systems (ISCAS)},
  pages={361--364},
  year={2014},
  organization={IEEE}
}

@article{ALLISON2019112,
title = {Design and Performance of an Interferometric Trigger Array for Radio Detection of High-Energy Neutrinos},
journal = {Nuclear Instruments and Methods in Physics Research Section A: Accelerators, Spectrometers, Detectors and Associated Equipment},
volume = {930},
pages = {112-125},
year = {2019},
issn = {0168-9002},
doi = {https://doi.org/10.1016/j.nima.2019.01.067},
url = {https://www.sciencedirect.com/science/article/pii/S016890021930124X},
author = {P. Allison and S. Archambault and R. Bard and J.J. Beatty and M. Beheler-Amass and D.Z. Besson and M. Beydler and M. Bogdan and C.-C. Chen and C.-H. Chen and P. Chen and B.A. Clark and A. Clough and A. Connolly and L. Cremonesi and J. Davies and C. Deaconu and M.A. DuVernois and E. Friedman and J. Hanson and K. Hanson and J. Haugen and K.D. Hoffman and B. Hokanson-Fasig and E. Hong and S.-Y. Hsu and L. Hu and J.-J. Huang and M.-H. Huang and K. Hughes and A. Ishihara and A. Karle and J.L. Kelley and R. Khandelwal and M. Kim and I. Kravchenko and J. Kruse and K. Kurusu and H. Landsman and U.A. Latif and A. Laundrie and C.-J. Li and T.C. Liu and M.-Y. Lu and A. Ludwig and K. Mase and T. Meures and J. Nam and R.J. Nichol and G. Nir and E. Oberla and A. ÓMurchadha and Y. Pan and C. Pfendner and M. Ransom and K. Ratzlaff and J. Roth and P. Sandstrom and D. Seckel and Y.-S. Shiao and A. Shultz and D. Smith and M. Song and M. Sullivan and J. Touart and A.G. Vieregg and M.-Z. Wang and S.-H. Wang and K. Wei and S.A. Wissel and S. Yoshida and R. Young},
collaboration = {ARA}
}

@article{PhysRevD.99.122001,
  title = {Constraints on the Ultrahigh-Energy Cosmic Neutrino Flux from the Fourth Flight of {ANITA}},
  author = {Gorham, P. W. and Allison, P. and Banerjee, O. and Batten, L. and Beatty, J. J. and Belov, K. and Besson, D. Z. and Binns, W. R. and Bugaev, V. and Cao, P. and Chen, C. H. and Chen, P. and Chen, Y. and Clem, J. M. and Connolly, A. and Cremonesi, L. and Dailey, B. and Deaconu, C. and Dowkontt, P. F. and Fox, B. D. and Gordon, J. W. H. and Hast, C. and Hill, B. and Hsu, S. Y. and Huang, J. J. and Hughes, K. and Hupe, R. and Israel, M. H. and Liewer, K. M. and Liu, T. C. and Ludwig, A. B. and Macchiarulo, L. and Matsuno, S. and McBride, K. and Miki, C. and Mulrey, K. and Nam, J. and Naudet, C. and Nichol, R. J. and Novikov, A. and Oberla, E. and Prohira, S. and Rauch, B. F. and Ripa, J. and Roberts, J. M. and Romero-Wolf, A. and Rotter, B. and Russell, J. W. and Saltzberg, D. and Seckel, D. and Schoorlemmer, H. and Shiao, J. and Stafford, S. and Stockham, J. and Stockham, M. and Strutt, B. and Sutherland, M. S. and Varner, G. S. and Vieregg, A. G. and Wang, N. and Wang, S. H. and Wissel, S. A.},
  collaboration = {ANITA Collaboration},
  journal = {Phys. Rev. D},
  volume = {99},
  issue = {12},
  pages = {122001},
  numpages = {11},
  year = {2019},
  month = {Jun},
  publisher = {American Physical Society},
  doi = {10.1103/PhysRevD.99.122001},
  url = {https://link.aps.org/doi/10.1103/PhysRevD.99.122001}
}
